\documentclass{article}

\usepackage{PRIMEarxiv}

\usepackage[utf8]{inputenc} %
\usepackage[T1]{fontenc}    %
\usepackage{hyperref}       %
\usepackage{url}            %
\usepackage{booktabs}       %
\usepackage{amsfonts}       %
\usepackage{nicefrac}       %
\usepackage{microtype}      %
\usepackage{lipsum}
\usepackage{fancyhdr}       %
\usepackage{graphicx}       %
\graphicspath{{media/}}     %

\usepackage{times}
\usepackage{soul}
\usepackage{graphicx}
\usepackage{amsmath}
\usepackage{amsthm}
\usepackage{algorithm}
\usepackage{algorithmic}
\usepackage{tabularx}
\usepackage{bm}
\usepackage{makecell}
\usepackage[figuresright]{rotating}
\usepackage{array}
\usepackage{colortbl}
\usepackage{multirow}
\newcolumntype{C}{>{\raggedright\arraybackslash}X}

\newcommand{\Description}[2][]{}

\title{Human Behavior Simulation: Objectives, Methodologies, and Open Problems
}

\author{
    Guozhen Zhang\thanks{Both authors contributed equally to this research.} \\
    TsingRoc \\
    Beijing, China \\
    \And
    Zihan Yu\footnotemark[1],
    Nian Li,
    Fudan Yu,
    Qingyue Long,
    Depeng Jin,
    Yong Li\thanks{Corresponding author. Email: liyong07@tsinghua.edu.cn} \\
    Department of Electronic Engineering, BNRist\\
    Tsinghua University \\
    Beijing, China \\
}

\begin{document}
\maketitle

\begin{abstract}
In recent years, human behavior simulation has drawn increasing attention from both academia and industry. The reasons fall into two aspects. First, simulation serves as a critical tool for understanding human behaviors, which has become one of the most important research topics in the history. Second, researchers have gradually reached a consensus that simulation, especially human behavior simulation, is critical for real-world decision-making systems. As a result, lots of human behavior simulation research and applications have sprung up across numerous disciplines in the past few years. In addition to the traditional methods, such as building mathematical and physical models, leveraging the recent advances of deep learning techniques -- especially the nascent Large Language Model technology -- for accurate human behavior simulation has also been one of the hottest research topics.  In this study, we provide a comprehensive review of the latest research advancements in human behavior simulation. We summarize the objectives, problem formulations, and commonly used methods and discuss the consistency in the development of related research in different disciplines, which reveals the gaps and opportunities for high-impact research in this promising direction. 
\end{abstract}

\keywords{Human Behavior \and Computer Simulation \and Artificial Intelligence}

\section{Introduction}

Simulation is typically referred to as a computational method that predicts the behavior or the outcome of a system~\cite{banks1999introduction,bratley2011guide}. It has greatly promoted the development of a wide range of disciplines, including physics~\cite{hollingsworth2018molecular,oberkampf2002verification}, biology~\cite{dada2011multi,dror2012biomolecular}, engineering~\cite{collins2021review}, climatology~\cite{roberts2018benefits,ravuri2021skilful}, etc., and supported many industrial applications, such as aircraft design~\cite{wang2014high}, drug discovery~\cite{durrant2011molecular}, and building design~\cite{attia2012simulation}. On the one hand, researchers use simulation as a tool to study complex systems and procedures, especially those in rare or extreme conditions. For example, they leverage biogeochemical cycle simulation to understand climate variability and change~\cite{hurrell2013community}. On the other hand, simulation creates a cost-effective and safe environment for decision-making. For example, researchers design materials for energy harvesting and storage based on molecular dynamic simulation~\cite{catlow2013computational} so that the design process is faster and requires fewer experiments. With the developments of data-driven algorithms, it also becomes the testbed for intelligent decision-making algorithms. For example, AlphaGo Zero is trained with a Go simulator~\cite{silver2017mastering}.

Traditionally, the simulation targets in most studies and applications are non-human objects, such as robotic arms and water molecules. In recent years, researchers and practitioners have gradually shifted their focus from objects to people~\cite{yang2020review,lazer2020computational}. In particular, the simulation of human behaviors has received a lot of attention, and the reasons come from two aspects. First, scientific research is increasingly concerned with understanding human behaviors to create a better life and society~\cite{tomavsev2020ai}. Second, in real-world decision-making applications, human behaviors often constitute the most important part of the environment. Accurate human behavior simulation is critical for building an effective decision-making environment. Further, human behaviors are highly complex, which poses a significant challenge to accurate simulation. The collection of a massive amount of behavioral data from the internet in recent years makes it possible for us to conduct large-scale data analysis and leverage recent advances in artificial intelligence for accurate human behavior simulation~\cite{lavin2021simulation}, the recently developed Large Language Model (LLM) technology also endows the capability to simulate human's complex behaviors, such as group actions~\cite{gao2023large,gao2023s,wang2023unleashing,park2023generative}, cognitive reasoning~\cite{wang2023unleashing,shah2022robotic,shah2023gnm}, and decision-making~\cite{gao2023large,gao2023s,shah2022robotic,shah2023gnm,zhu2023ghost,shinn2023reflexion}. As a result, lots of new research and applications have sprung up across numerous disciplines.

We seek to bring these newest advancements in human behavior simulation across disciplines together and provide a systematic introduction to the motivations, tasks, challenges, and methods for the research community, which reveals the gaps and opportunities for high-impact research in this promising direction. We summarize the consistency in the development of related research in different disciplines and hope that scientists obtain insights from the research in different fields so that they can study novel problems, apply techniques to problems that they were not designed for, and explore new directions of human behavior simulation.

Compared with existing surveys on simulation, this work takes a brand new perspective --- distinguishing works from their simulation targets. Existing surveys either take a view of simulation methods, such as a review of agent-based modeling~\cite{wall2016agent,bianchi2015agent}, or a view of research disciplines, such as molecular dynamics simulation~\cite{hollingsworth2018molecular} or supply chain simulation~\cite{oliveira2016perspectives}. However, many commonalities exist in the simulation of different human behaviors, and a comprehensive survey is lacking. This work fills this gap, and we can summarize our contribution as four-fold. First, to the best of our knowledge, this paper presents an updated and comprehensive survey on recent advances in human behavior simulation for the first time. Second, we summarize the problems and main challenges for human behavior simulation and organize them into an appropriate taxonomy. Third, we discussed the common and the most advanced methods, for example, artificial intelligence and the large language model, for human behavior simulation. Finally, we highlight open problems and challenges to facilitate future research on this promising topic.

The rest of the survey is organized as follows: Section 2 gives a brief introduction to human behaviors and their taxonomy. Section 3 describes the common objectives pursued through human behavior simulation and presents the latest research problems of human behavior simulation from different research disciplines. Section 4 discusses the main challenges and summarizes the common methods for human behavior simulation. Section 5 discusses the open problems, challenges, and possible future directions. Section 6 contains concluding remarks.

\section{A Short Primer on Human Behavior}

Human behavior is the activity of individuals or groups to meet their needs and adapt to the environment~\cite{skinner1965science}. It results from genetic, physiological, psychological, social, and other internal and external factors and can also be seen as a product of human interaction and the environment. Considering the origin, performance, and characteristics of behavior, it can be broadly classified into four categories: cognitive behavior~\cite{tomasello2003makes}, physiological behavior~\cite{cooke2014physiology}, economic behavior~\cite{neff2017work}, and social behavior~\cite{enfield2006roots}. 

\begin{figure}[t]
    \centering
    \includegraphics[width=0.5\linewidth]{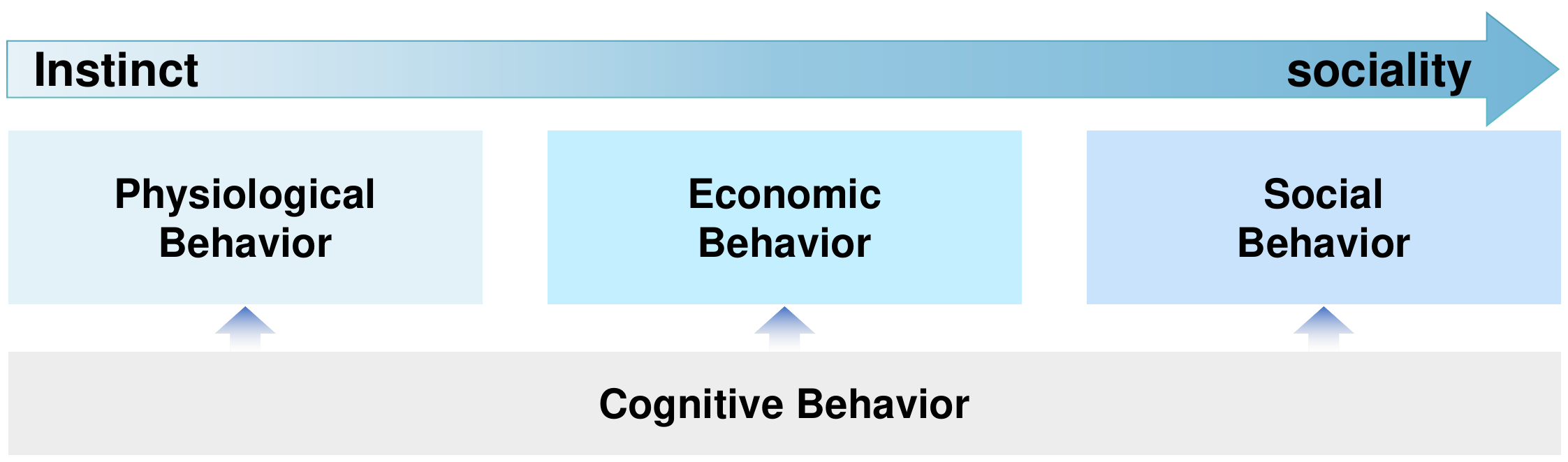}
    \caption{The relationship between the four types of behavior.}
    \label{fig:relationship}
    \Description[The relationship between the four types of behavior.]{The relationship between the four types of behavior.}
\end{figure}

\begin{itemize}
  \item \textbf{Cognitive Behavior.} Cognitive behavior refers to humans' common mental processes, including reasoning, creativity, the experience of emotion, and politico-religious behavior~\cite{evans1993human,evans2003emotion,turner2006artful,driskell2008faith}. Specifically, reasoning is the ability to summarize and generalize experience and apply it to new and unknown situations. Creativity refers to using previous ideas or resources to create something original. Emotions, such as pleasure, shame, and love, are an essential type of mental experience. Politico-religious behavior is a set of activities based on religious or political beliefs. These cognitive behaviors can spread from person to person via interactions and form complex phenomena in society.
  \item \textbf{Physiological Behavior.} Physiological behavior can be understood as actions that maintain our basic living needs. Some physiological behaviors are common to all animals, such as eating and sleeping. Humans further move and leverage resources, such as electricity and networks, to meet their living needs.
  \item \textbf{Social Behavior.} Social behavior refers to the interactions between two or more people, such as making friends, competition, and cooperation~\cite{snyder1985personality}. It also includes the effects of the interaction, such as the influence of one on another. Sometimes, interactions change not only people's behavior but also their minds.
  \item \textbf{Economic Behavior.} Economic behavior accounts for actions regarding the development, organization, and use of materials and other forms of work. Humans consider economic decisions through cost-benefit analysis and risk-reward considerations. The three most common economic behaviors are work, entertainment, and market behavior~\cite{teitelbaum2018research}. The complexity of society determines the nature of human work. In primitive societies, work was not considered a unique activity but a constant that constituted all parts of life. Gradually, as society developed, humans could perform more specialized work in their respective fields, i.e., non-manual work, with some specializing in technical knowledge and management. Then, mass production industries and services were gradually created. Nowadays, people deal with work differently depending on their physical and personal characteristics. Entertainment is an activity performed outside work that temporarily relieves psychological stress, generates positive emotions, or promotes social interactions. Market behavior refers to the behavior of customers, sellers, firms, and other market participants to maximize their own interests. In this survey, we focus on the behavior of individuals.
\end{itemize}

We present the relationship among cognitive, physiological, economic, and social behavior in Figure~\ref{fig:relationship}. Cognitive behavior is the foundation of the other three since human consciousness affects their actions~\cite{newell1994unified}. For example, one's emotion affects his/her consumption choices. Further, from another perspective, physical behavior is more instinctive than other behavior. In other words, this is a conditioned reflex inherited in our genes. Therefore, the heterogeneity and randomness between different people are relatively weak.
On the contrary, from physical behavior to economic behavior to social behavior, the sociality of behavior is getting stronger and stronger~\cite{lu2021does}. Different societies have different behavioral norms and cultures, and thereby social behavior and economic behavior are much more heterogeneous than physical behavior. As a result, these behaviors are also more complex and more difficult to simulate.

\section{Problems and Objectives}

\begin{figure}
    \centering
    \includegraphics[width=0.5\linewidth]{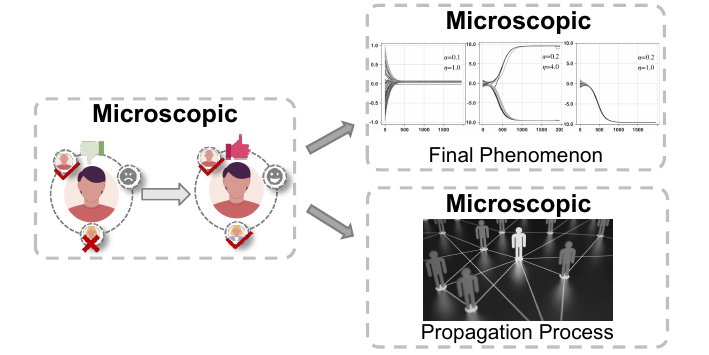}
    \caption{The illustration of cognitive behavior simulation.}
    \label{fig:cognitive}
    \Description[The illustration of cognitive behavior simulation.]{The illustration of cognitive behavior simulation.}
\end{figure}

Recently, a wide range of disciplines are paying increasing attention to human behavior simulation. Although different disciplines formulate human behavior simulation problems into different specific forms, their objectives can be broadly summarized into two categories:

\begin{table*}[htbp]
    \centering
    \begin{tabular}{clp{10em}p{10em}}
\toprule
\multicolumn{1}{p{3em}}{ } &                           & \multicolumn{2}{p{20em}}{\textbf{for Scientific Discovery}} \\
\cmidrule{3-4}\multicolumn{1}{p{3em}}{ } &                           & Microscopic               & Macroscopic \\
\midrule
\multicolumn{1}{c}{\multirow{3}[6]{*}{\textbf{Cognitive}}} & \cellcolor[rgb]{ .929,  .929,  .929}Emotion & \cellcolor[rgb]{ .929,  .953,  .859}\cite{bilewicz2020hate} & \cellcolor[rgb]{ .729,  .82,  .443}\cite{lakon2017cascades,burns2007diffusion,mathew2019spread} \\
\cmidrule{2-4}                          & \cellcolor[rgb]{ .929,  .929,  .929}Creativity & \cellcolor[rgb]{ .929,  .953,  .859}\cite{rabinovich2020sequential} &   \\
\cmidrule{2-4}                          & \cellcolor[rgb]{ .929,  .929,  .929}Politico-Religious Behavior & \cellcolor[rgb]{ .929,  .953,  .859}\cite{rossman2021network} & \cellcolor[rgb]{ .729,  .82,  .443}\cite{xie2021detecting,dykstra2013put,cinelli2021echo} \\
\midrule
\multicolumn{1}{c}{\multirow{4}[8]{*}{\textbf{Physiological}}} & \cellcolor[rgb]{ .886,  .949,  .973}Physical Movement & \cellcolor[rgb]{ .729,  .82,  .443}\cite{helbing1995social,XUE2020bidirecionflow,moussaid2011simple} & \cellcolor[rgb]{ .729,  .82,  .443}\cite{hughes2002continuum,bellomo2008modelling,jiang2010higher,maury2010macroscopic,twarogowska2014macroscopic} \\
\cmidrule{2-4}                          & \cellcolor[rgb]{ .886,  .949,  .973}Driving & \cellcolor[rgb]{ .729,  .82,  .443}\cite{sun2019modelling,li2021modified,liu2020modeling,Lin2019CFauto,Tan2022CFrisk,Park2020ACF,Ro2018CFauto,Lindorfer2018imperfect,Shang2022ACF,9901459,li2022gametheoretic} &   \\
\cmidrule{2-4}                          & \cellcolor[rgb]{ .886,  .949,  .973}Trajectory-based Transition & \cellcolor[rgb]{ .929,  .953,  .859}\cite{Jiang2016EPR} & \cellcolor[rgb]{ .929,  .953,  .859}\cite{Levin2021-ge,Ferreira2018Scale} \\
\cmidrule{2-4}                          & \cellcolor[rgb]{ .886,  .949,  .973}Resource Usage &                           & \cellcolor[rgb]{ .929,  .953,  .859}\cite{Yoon2020pnas,LI2019451} \\
\midrule
\multicolumn{1}{c}{\multirow{3}[6]{*}{\textbf{Social}}} & \cellcolor[rgb]{ .769,  .937,  1}Social Connection Formation & \cellcolor[rgb]{ .929,  .953,  .859} \cite{barabasi1999emergence} & \cellcolor[rgb]{ .729,  .82,  .443}\cite{asikainen2020cumulative,bianconi2014triadic} \\
\cmidrule{2-4}                          & \cellcolor[rgb]{ .769,  .937,  1}Social Influence & \cellcolor[rgb]{ .729,  .82,  .443}\cite{schelling1971dynamic,laurie2003role,zhang2004residential} & \multicolumn{1}{c}{} \\
\cmidrule{2-4}                          & \cellcolor[rgb]{ .769,  .937,  1}Cooperation and Competition &                           & \cellcolor[rgb]{ .729,  .82,  .443}\cite{cohen2001role,nowak1992tit,boero2010bother,conte2002reputation,gintis2000strong,bowles2004evolution,macy1998evolution,bravo2012trust,fehl2011co} \\
\midrule
\multicolumn{1}{c}{\multirow{3}[6]{*}{\textbf{Economic}}} & \cellcolor[rgb]{ .784,  .89,  .984}Work &                           & \multicolumn{1}{c}{} \\
\cmidrule{2-4}                          & \cellcolor[rgb]{ .784,  .89,  .984}Entertainment &                           &   \\
\cmidrule{2-4}                          & \cellcolor[rgb]{ .784,  .89,  .984}Market & \cellcolor[rgb]{ .729,  .82,  .443}\cite{gode1993allocative,palmer1994artificial,cliff1998less,albin1992decentralized} & \cellcolor[rgb]{ .729,  .82,  .443}\cite{farmer2009economy,byrd2020abides,karpe2020multi,amrouni2022ctmstou} \\
\bottomrule
\end{tabular}%

    \caption{The research progress on the \emph{scientific discovery} objective in different research fields. Here, the shades of color in the table represent the number of related works. The darker the color, the more related works. The same as Table~\ref{tab:forDM}.}
    \label{tab:forSD}
\end{table*}

\begin{table*}[htbp]
    \centering
    \begin{tabular}{ccp{10em}p{10em}}
\toprule
\multicolumn{1}{p{3em}}{ } &                           & \multicolumn{2}{p{20em}}{\textbf{for Decision-Making}} \\
\cmidrule{3-4}\multicolumn{1}{p{3em}}{ } &                           & Microscopic               & Macroscopic \\
\midrule
\multicolumn{1}{c}{\multirow{3}[6]{*}{\textbf{Cognitive}}} & \cellcolor[rgb]{ .929,  .929,  .929}Emotion &                           & \cellcolor[rgb]{ .929,  .953,  .859}\cite{lopez2021simulating} \\
\cmidrule{2-4}                          & \cellcolor[rgb]{ .929,  .929,  .929}Creativity &                           &   \\
\cmidrule{2-4}                          & \cellcolor[rgb]{ .929,  .929,  .929}Politico-Religious Behavior & \cellcolor[rgb]{ .929,  .953,  .859}\cite{zhou2020two,chen2021opinion} & \cellcolor[rgb]{ .929,  .953,  .859}\cite{santos2021link,baumann2020modeling} \\
\midrule
\multicolumn{1}{c}{\multirow{4}[8]{*}{\textbf{Physiological}}} & \cellcolor[rgb]{ .886,  .949,  .973}Physical Movement & \cellcolor[rgb]{ .729,  .82,  .443}\cite{helbing2000simulating,alahi2016socialLSTM,gupta2018social,mohamed2020socialSTGCNN,zhang2022physics,yu2023understanding} &   \\
\cmidrule{2-4}                          & \cellcolor[rgb]{ .886,  .949,  .973}Driving & \cellcolor[rgb]{ .729,  .82,  .443}\cite{SAPRYKIN2019199,astarita2019traffic,Suo_2021_CVPR,peng2022learning,zhang2022systematic,zhou2022short,zhu2021combined,9088249,Chu2021Individual,zhang2022trajgen,wei2022game,9580542} &   \\
\cmidrule{2-4}                          & \cellcolor[rgb]{ .886,  .949,  .973}Trajectory-based Transition & \cellcolor[rgb]{ .729,  .82,  .443}\cite{Luo2022RLMob,Feng2020Simulate,yuan2022activity,jiang2018deep,Fan2018Online,Feng2020PMF,Feng2018DeepMove} & \cellcolor[rgb]{ .729,  .82,  .443}\cite{Wang2019OD,Shi2020OD,Ruan2020Allocation,Wu2018Planning} \\
\cmidrule{2-4}                          & \cellcolor[rgb]{ .886,  .949,  .973}Resource Usage & \cellcolor[rgb]{ .729,  .82,  .443}\cite{Mahmood2020SIMULATION,schumann:hal-03505242,Xu2021STAN,RING2019156,Hui2022Knowledge} & \cellcolor[rgb]{ .729,  .82,  .443}\cite{w12071885,w12061628,8790060,Mahmood2020SIMULATION,8057090,8466626,8667446} \\
\midrule
\multicolumn{1}{c}{\multirow{3}[6]{*}{\textbf{Social}}} & \cellcolor[rgb]{ .769,  .937,  1}Social Connection Formation & \cellcolor[rgb]{ .729,  .82,  .443}\cite{geng2015learning,qu2020continuous,pareja2020evolvegcn,sankar2020dysat,zhou2018dynamic,wang2023neural} & \multicolumn{1}{c}{} \\
\cmidrule{2-4}                          & \cellcolor[rgb]{ .769,  .937,  1}Social Influence & \cellcolor[rgb]{ .929,  .953,  .859}\cite{lu2018collective,cinus2020generating} & \cellcolor[rgb]{ .929,  .953,  .859}\cite{granovetter1978threshold,hedstrom1994contagious} \\
\cmidrule{2-4}                          & \cellcolor[rgb]{ .769,  .937,  1}Cooperation and Competition &                           &   \\
\midrule
\multicolumn{1}{c}{\multirow{3}[6]{*}{\textbf{Economic}}} & \cellcolor[rgb]{ .784,  .89,  .984}Work & \cellcolor[rgb]{ .729,  .82,  .443}\cite{rachid2018agent,somarathna2020agent,tian2022hybrid,ajmi2019agent} & \cellcolor[rgb]{ .729,  .82,  .443}\cite{baskaran2019digital,troost2020bioeconomic,troost2020bioeconomic,chaudhari2018putting,firdausiyah2019modeling,ruvzic2021simulation,chen2017crowdsourced} \\
\cmidrule{2-4}                          & \cellcolor[rgb]{ .784,  .89,  .984}Entertainment & \cellcolor[rgb]{ .729,  .82,  .443}\cite{shi2019virtual,luo2022mindsim,goodfellow2020generative,zhao2021usersim,zhang2020evaluating} & \multicolumn{1}{c}{} \\
\cmidrule{2-4}                          & \cellcolor[rgb]{ .784,  .89,  .984}Market & \cellcolor[rgb]{ .729,  .82,  .443}\cite{nowak1992evolutionary,aktipis2004know,aktipis2006recognition,aktipis2011cooperation}  & \cellcolor[rgb]{ .729,  .82,  .443}\cite{mi2023taxai,zheng2020ai,curry2022analyzing} \\
\bottomrule
\end{tabular}%

    \caption{The research progress on the \emph{decision-making} objective in different research fields.}
    \label{tab:forDM}
\end{table*}

\begin{itemize}
    \item Simulation as a scientific tool for analyzing human behavior. On the one hand, researchers use simulation as an analytical tool for scenarios with high complexity. Taking queuing theory as an example, traditional mathematical methods fail to give analytic solutions for average waiting time when the service time distribution is complex with multiple cashiers. On the contrary, simulation can effectively derive the results under any complexity. On the other hand, simulation is introduced as an effective method to discover the laws and mechanisms of human behavior. In general, this type of research focuses on the interpretability of the simulation model.
    \item Simulation for building a decision-making environment. With more and more research on decision problems in real-world scenarios with complex constraints and unclear mechanisms, researchers have reached a consensus that simulation is critical for decision-making. Effective simulation enables us to deduce how things develop in different conditions and answer counterfactual questions. In real-world scenarios, human behavior simulation is the most important and difficult part of building an effective decision-making environment. Thus, how to simulate human behavior accurately has become the goal of many studies. 
\end{itemize}

In this section, we present the latest research on human behavior simulation from different research disciplines together with their research objectives. We summarize them in Table~\ref{tab:forSD} and Table~\ref{tab:forDM}, which clearly shows that both objectives receive lots of attention, and different research disciplines have different focuses and progress. Further, existing studies on human behavior simulation have two perspectives: microscopic perspective and macroscopic perspective. Works from the microscopic perspective focus on the behavior of each individual, while those from the macroscopic perspective concentrate on the behavior of groups and communities. In this section, we aim to provide a systematic view of the research progress to guide the researchers in different disciplines for future research. Note that we do not intend to be exhaustive in this survey, and thus those behaviors with few works simulating it are not included in this section.

\subsection{Cognitive Behavior}

Existing work on cognitive behavior simulation has focused primarily on simulations of emotion, creativity, and politico-religious behavior. Most existing work does not focus on how these cognitive processes are formed in our minds but rather concentrates on simulating how these cognitive behaviors evolve and how they spread between people, and further explores what phenomena are ultimately caused by the cognition of individuals in a macroscopic society. Specifically, cognitive behavior refers to the modeling of the change or propagation process of cognition, described as follows:
\begin{equation}
{x_{i}(t+1)=F\left(x_{i}(t),\left\{x_{j}(t)\right\}_{j \in N_{i}}, \theta\right)}
\end{equation}
where $x_{i}(t)$ represents the cognition (such as opinion, emotion) status of the individual $i$ at time $t$, $x_{i}(t+1)$ is the cognition of the individual $i$ at the next time step. $F$ is a function that describes how cognition is updated, which can depend on various factors including the current cognition status of the individual, the set of opinions of its interacting neighbors $\{x_{j}(t)\}_{j \in N_{i}}$, and other possible parameters $\theta$ such as trust thresholds or influence weights.

\subsubsection{Emotion}
Researchers mainly utilize simulation to explore the mechanisms of diffusion and propagation of various emotions, such as love~\cite{lakon2017cascades}, fear~\cite{burns2007diffusion}, and hate~\cite{mathew2019spread}. Since these researches are generally similar, we choose a typical example, the simulation of hate, to demonstrate their problems and objectives.

Researchers suggest that adequate simulation helps us understand the fundamental law of its generation and impact. A standard research paradigm in this field is to construct interpretable models drawing on empirical knowledge from multiple disciplines.
For example, Bilewicz et al.~\cite{bilewicz2020hate} focus on how hate speech ultimately affects intergroup relations and whether it causes political radicalization. Specifically, using empirical evidence from social and emotional psychology, they propose a simulation model to explain how exposure to hate speech affects people's emotional, behavioral, and normative dimensions.

In addition to discovering the mechanisms behind hate and its corresponding behaviors, researchers also leverage simulation for better decision-making. For example, Lopez et al.~\cite{lopez2021simulating} focus on how hate speech spreads between individuals, and simulating such a spreading process is beneficial for the government in formulating policies to regulate hate speech. Another typical example is $S^3$~\cite{gao2023s}, which employs agents powered by Large Language Models (LLMs) to imitate both individual and group actions in a social network setting. $S^3$ adeptly duplicates human characteristics like emotional responses, attitudes, and interactive patterns. It utilizes data from actual social networks to create its simulation framework, in which information shapes the emotions and subsequent actions of the users.

\subsubsection{Creativity}
Simulation can be used to discover how creativity is formed and to explain the mechanisms by which it arises. For example, Rabinovich et al.~\cite{rabinovich2020sequential} establish brain information processes' sequential dynamics through coupled heteroclinic networks and metastable states. They described the mechanism of creativity generation in the brain using a set of nonlinear differential equations solved in terms of transferable patterns in various cognitive processes, which successfully explores the mechanisms of creativity formation and development. Wang et al. ~\cite{wang2023unleashing} developed Solo Performance Prompting (SPP), a method that mimics human cognitive collaboration by turning a single Large Language Model (LLM) into a multi-faceted agent. This approach improves the solving of complex reasoning and specialized knowledge tasks, and greatly surpasses traditional techniques in activities like trivia writing and logic puzzles, demonstrating its strength in joint problem-solving.

\subsubsection{Politico-religious Behavior}
Most studies on simulating politico-religious behavior focus on how people's opinions propagate and change~\cite{rossman2021network,dykstra2013put,cinelli2021echo}.

Regarding propagation simulation, the main goal of existing works is to use simulation as a tool to study, analyze, and understand human cognitive behavior and to provide a high level of interpretability. For example, some researchers reproduce the essential changes in the diffusion rate and mutation rate of information propagation on social media at a macro level, specifically using the theory of percolation phase transition from physics to simulate this process~\cite{xie2021detecting}. Subsequent studies have further focused on the impression of external factors on diffusion (i.e., considering the external influences of advertising and mass media), improving the traditional Bass diffusion model to incorporate these factors into the modeling~\cite{rossman2021network}.

In terms of opinion evolution simulation, most studies have focused on explaining phenomena and patterns of cognitive behavior using simple and interpretable simulation models.
For instance, some studies focus on the macro phenomena during evolving opinion, such as whether opinions will become neutral, bipolar, or unipolar over time. Specifically, these methods model how the topology of social networks evolves as well as how this evolution affects opinion dynamics, and further simulate how people's opinions evolve from initial opinions to extreme opinions~\cite{santos2021link,baumann2020modeling}.
Some studies further focus on the impact of certain special individuals on the overall change of opinions. Specifically, they introduced hard-to-persuade or biased individuals into the model design and developed a more comprehensive and accurate simulator with stronger interpretability, which helps to make scientific findings and can be further used for decision-making~\cite{zhou2020two,chen2021opinion}. %

\subsection{Physiological Behavior}
Physiological behavior simulation is the hottest and most developed one among all types of behavior and has been widely researched in recent years. The most mature research direction is simulating how people move~\cite{dong2019state,kulkarni201920}. According to the simulation granularity of time and space, it can be divided into three categories: 1) \emph{Physical Movement}, which simulates how people walk and run, 2) \emph{Driving}, which simulates how people drive vehicles, and 3) \emph{Transition}, which studies the coarse-grained trajectory of people moving from one place to another in their daily lives. 
Researchers leverage simulation to understand these behaviors and further facilitate various decision-making problems, such as building architecture design~\cite{feng2016crowd}, traffic management~\cite{otsuka2019bayesian}, and emergency planning~\cite{shendarkar2006crowd}.

Another rising research direction in physiological behavior simulation is simulating the usage of resources, such as electricity and network~\cite{LI2019451}, which can be seen as human activity's footprint. Existing works mainly utilize resource usage simulation for better resource management and control of carbon emissions~\cite{schumann:hal-03505242,masciadri2018disseminating}.

\subsubsection{Physical Movement}

\begin{figure}
    \centering
    \includegraphics[width=0.5\linewidth]{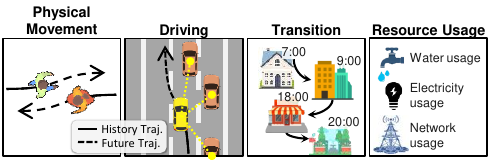}
    \caption{The illustration of Physiological behavior simulation.}
    \label{fig:physiological}
    \Description[The illustration of Physiological behavior simulation.]{The illustration of Physiological behavior simulation.}
\end{figure}

Physical movement simulation aims to simulate how pedestrians move and interact with each other in certain places, such as airport waiting rooms and traffic intersections~\cite{LI2021platform,Song2019exit}.

Physical movement simulation was originally used to understand and prevent crowd disasters, such as crowd collapses and crowd crushes\cite{yang2020review}. These researches usually focus on scenarios with overcrowded people. In these scenarios, people are so close to each other that the crowd can be viewed as an incompressible flow, and the pedestrian movement is dominated by the squeezing of others rather than individual will. Under these settings, Hughes et al. \cite{hughes2002continuum} proposed a macroscopic physical movement model. They defined the crowd as a “thinking fluid”, and derived a set of partial differential equations based on a mass conservation law to describe the evolution of crowd density $\rho(\bm{p},t)$, where $\rho(\bm{p},t)$ represents the crowd density of position $\bm{p}$ at time $t$. This macroscopic modeling paradigm is inherited by some subsequent research, which redesigned the crowd density evolution equation through gas dynamics equations\cite{bellomo2008modelling,jiang2010higher}, gradient flow methods\cite{maury2010macroscopic}, and other hand-designed methods\cite{twarogowska2014macroscopic} to provide insights into the phenomena observed in dense crowds, such as stop-and-go waves and turbulent flows\cite{helbing2007dynamics}.

These years, as fields like traffic scheduling, architectural designing, and urban planning are placing an ever-growing emphasis on human movement simulation under normal conditions\cite{yang2020review}, and more importantly, with the increasing availability of individual movement trajectory data, microscopic models have progressively become the primary paradigm in the realm of physical movement modeling\cite{zhang2022physics}. Unlike macroscopic models that focus on the evolution of crowd density, these microscopic models focus on the movement trajectories of each individual. Formally, given the initial position $\bm{p}_i(0)$ and destination $\bm{d}_i$ of each pedestrian $i$, a microscopic model aims to simulate their future trajectories $\{\bm{p}_i(t)| t = 0, \Delta t, 2\Delta t, \cdots\}$, where $\Delta t$ is the time step of simulation.

Helbing\cite{helbing1995social} first proposed a microscopic model, known as the social force model (see Section~\ref{sec:SFM} for details), to simulate and explain the phenomena observed when pedestrians go through bottlenecks or corridors. Based on this model, they further developed a method\cite{helbing2000simulating} to investigate the impact of building shape and exit layout on evacuation times during emergencies, which can serve as a guide for optimizing building design and emergency evacuation strategies. In addition to forces, subsequent work has used cellular automata models\cite{XUE2020bidirecionflow}, utility fuctions\cite{moussaid2011simple}, and neural netowks\cite{zhang2022physics,yu2023understanding} to simulate more phenomena observed in the real world and thus realize more realistic simulation. 

Corresponding to the thinking process during human movement, some other works also utilized the recently emerged Large Language Models (LLMs) to simulate human movement trajectories. For example, Shah et al.\cite{shah2022robotic} introduced an LM-Nav system, which can generate navigation instructions based on given observations in a target environment with a pre-trained LLM. Following this, they further\cite{shah2023gnm} developed a goal-conditioned model to simulate human-like vision-based movement in complex physical environments.

Other fields such as computer vision and autonomous driving are also beginning to focus on human movement behavior, but they focus on different objectives. In the realm of computer vision, the primary concern for researchers is not necessarily achieving perfect realness, but rather ensuring the pedestrian can efficiently move to its destination along a collision-free path. Therefore, they often employ rule-based methods. For example, Lee et al.\cite{lee2018crowd} and Hu et al.\cite{hu2021heterogeneous} have leveraged reinforcement learning to optimize rule-based reward functions, which offer rewards for agents successfully reaching their destinations, and impose penalties when they collide with others. In contrast, in the domain of autonomous driving, the main objective revolves around predicting pedestrian movement to enhance the safety of autonomous vehicles. Consequently, these works typically prioritize prediction accuracy (i.e., how similar the predicted trajectory is to the real trajectory) rather than modeling pedestrian interactions, despite the latter being a central aspect of conventional human movement simulation work. Therefore, works in this field\cite{alahi2016socialLSTM,gupta2018social,mohamed2020socialSTGCNN} usually adopt supervised learning and focus on generating trajectories as close to real-world trajectories as possible.

\subsubsection{Driving}

As driving is one of the most common ways people travel nowadays, the simulation of driving behavior is of great significance for many disciplines and applications, such as traffic management, intelligent transportation systems, and autonomous driving. Simulation is broadly used for 1) understanding drivers' driving processes and whole traffic dynamics, 2) helping traffic management by analyzing how drivers' behaviors affect traffic dynamics, and 3) improving autonomous decision-making by providing a realistic environment for its training and testing. Existing works can be divided into three categories by behaviors they simulated, including longitudinal behavior, transverse behavior, and collision-avoiding behavior. The studied problems can be generally formulated as predicting the position and speed of a target vehicle $i$ given the current and historical positions and speeds of the target vehicle and its neighboring vehicles $N_i$:
\begin{equation}
\begin{split}
x(t+1),v(t+1) = F( x(t-T:t],v(t-T:t], \\ \{x_{i}(t-T:t]\}_{1 \leq i \leq N}, \{v_{i}(t-T:t]\}_{1 \leq i \leq N} ),
\end{split}\label{eq:driving}
\end{equation}
where the $x(t-T:t]$ and $v(t-T:t]$ are the historical position and speed of the target vehicle during $T$ time steps, and $\{x_{i}(t-T:t]\}_{1 \leq i \leq N}$, and $\{v_{i}(t-T:t]\}_{1 \leq i \leq N}$ are the historical position and speed of neighboring vehicles.

Simulating the longitudinal driving behavior, i.e., the driver's car-following behavior, is to model the acceleration and deceleration decisions under the scenario that the driver safely controls a following vehicle behind a leading vehicle within the same lane while avoiding collision. In car-following models, the scenario is often one-dimensional with a single driving lane, so only the longitudinal positions are considered in Equation \ref{eq:driving}. Besides, the neighboring vehicles are often restricted to the single leading vehicle that drives in front of the target following vehicle. 
Traditionally, researchers derive mathematical formulas as rule-based models from safety constraints~\cite{ZHANG2021review}. Representative models include Gazis-Herman-Rothery model~\cite{Denos1961GHR}, Gipps model~\cite{GIPPS1981105}, intelligent driver model (IDM)~\cite{IDM} and Wiedemann model~\cite{Wiedemann}. For example, the IDM model ~\cite{IDM} assumes a principle that the following vehicle maintains enough distance from the preceding vehicle to prevent a collision to derive the governing equations of the safety distance and acceleration. Recent works focus more on modeling human driving characteristics rather than only physical safety constraints, with the observation that traditional mathematical models tend to produce overly safe driving traces. For example, recent works attempts to model reaction delay~\cite{Lindorfer2018imperfect}, distraction and errors~\cite{Ro2018CFauto}, personal styles~\cite{Chu2021Individual}, the asymmetric nature of human behavior~\cite{Park2020ACF,Shang2022ACF}, the memory effect and prediction capacities of human drivers~\cite{Wang2018deep}, and the anticipation of risk~\cite{Tan2022CFrisk}. 

Transverse driving behavior is about the driver's lane-changing behaviors. In lane-changing processes, both the longitudinal and lateral positions of vehicles are considered in Equation \ref{eq:driving}, and thus the neighboring vehicles can be extended to at most 6 nearest vehicles within the perception distance of the targeted driver, namely, the front and the behind vehicle at the same lane, as well as the left and the right lane. The preliminary output of the model can be a probability distribution over the 3 possible lane-changing decisions, namely, keeping the lane, changing to the left lane, and changing to the right lane, based on the estimation of the perceived lane-changing utility and risks. Finally, the model can further combine the lane-changing decisions with car-following models to execute the lane-changing process and finally derive the prediction of the longitudinal and lateral position and speed of the targeted vehicle at the next time step.
For instance, Liu et al.~\cite{liu2020modeling} derive an overtaking flow chart to model the choice of the driver based on the heterogeneous speeds between vehicles, the safe space for lateral operations, and the utility for overtaking. Zhou et al.~\cite{zhou2022short} calculate a short-term preview of the transverse movement of target vehicles and further predict the lane-changing actions based on probabilistic models. Zhu et al.~\cite{zhu2021combined} combines physical safety mechanisms with data-driven techniques to arrive at models that can both generally capture the lane-changing dynamics and fit into personal styles.

Research focusing on collision-avoiding behavior puts special interest on the complex competitive and cooperative behaviors such as cutting-in, rushing, yielding, and waiting behaviors, which frequently happen at those spots that are especially prone to collision rather than the normal driving lanes, such as intersection and on-ramp junctions. Some works explicitly model the competitive and cooperative behaviors with the game theory, which can well capture the social interactions at collision-prone spots~\cite{Wilko2019Socialav,wei2022game,li2022gametheoretic}. Others leverage expert knowledge and data-driven methods, such as multi-agent reinforcement learning, to further improve simulation fidelity.

Some researchers involve all kinds of driving behaviors for comprehensive traffic simulation. Many traffic simulators are built for traffic management~\cite{9088249,SAPRYKIN2019199,Suo_2021_CVPR}. Recent works further improve the level of simulation realism by learning to model driving behaviors from human demonstrations with neural networks~\cite{zhang2022systematic,zhang2022trajgen} and to calibrate physical models with neural networks~\cite{9580542}.

As a new modeling paradigm, LLM has been explored to serve as the simulated drivers in driving scenarios. For example, LLM is found to be able to understand its target and historical environment state and further decide the actions such as speed up and change lane, which is shown to be human-like and safe\cite{jin2023surrealdriver}.

\subsubsection{Transition}
Transition simulation studies the coarse-grained trajectory of people moving from one place to another in their daily lives. To be specific, if we list the places one has been to in one day, we can get a coarse-grained trajectory, i.e., his transition trajectory. Formally, a transition trajectory is a sequence of spatial-temporal points:
\begin{equation}
S = [x_1, x_2, ..., x_n], \text{where } x_i = (l_i, t_i),
\end{equation}
and the location $l_i$ could be represented by the ID of the visited PoI (Point of Interest), AoI / RoI (Area / Region of Interest), or GPS coordinates, according to the studied scenario. 
Several works aim to use explainable parameters and mechanisms to model, analyze, and understand the transition~\cite{Jiang2016EPR,Ferreira2018Scale,Levin2021-ge}. In this case, a flow-chart-like mechanism is typically developed to model the human transition schedules. For example, Jiang et al.~\cite{Jiang2016EPR} implements an extended Exploration and Preferential Return (EPR) model to generate individual mobility trajectories based on the inferred daily life pattern from sparse observations. 
Recent years have witnessed more works aiming at generating or predicting the human transition trajectories utilizing data-driven methods rather than knowledge-based or physics-based models~\cite{jiang2018deep,Fan2018Online,Feng2018DeepMove,Feng2020PMF,Luo2022RLMob,Feng2020Simulate,yuan2022activity}, to effectively learn the complex high-order spatiotemporal dynamics of human mobility. The generation task aims to generate high-fidelity synthetic trajectories by learning the distribution of the collected real trajectory data, so that given a sample of real trajectories, large amounts of lifelike trajectories can be obtained for downstream applications such as traffic simulation. For instance, Feng et al.~\cite{Feng2020Simulate} proposes a GAN framework with encoded prior knowledge in urban structure by an attention-based region network to generate the trajectory. The prediction task aims to predict trajectory points by learning the conditional distribution on the historical trajectories, so that proper life service recommendations can be made given the prediction of the user's visitation intention, or public resources can be dynamically allocated given the predicted change of mobility~\cite{Ruan2020Allocation}. Some works formulate this problem as an aggregated origin-destination prediction problem, which predicts the flow number between a pair of locations can be applied in helping the planning of ride-hailing services~\cite{Wu2018Planning,Wang2019OD,Shi2020OD}. 
There are also pioneering works that utilize LLM to predict human transition trajectories, where the LLM is found to be able to learn the long-term patterns and short-term stochasticity of historical trajectories organized in tabular form and then predict the location and stay time of the next movement~\cite{wang2023would}. These works, together with the walking and driving simulation research, finally support an urban-scale traffic simulator to provide a realistic and cost-free environment for decision-making applications in urban service~\cite{zhang2022mirage}.

\subsubsection{Resource Usage}
Simulating the use of resources, such as water, electricity, and networks, has also been a rising research topic. Researchers leverage it to facilitate city infrastructure planning, resource allocation, conservation strategy designing, and carbon footprint tracking. The formulation of the problems can be generally summarized as predicting or generating the amount of used water, electricity, or network traffic by individuals or a group of humans in a given region such as a building or a city with a resolution from daily dynamics to yearly estimations.

In terms of water usage simulation, many works adopt recently developed data-driven methods to predict water demand based on rich historical data. Here the human behavioral factors are not explicitly modeled while the results of the water usage, namely the consumed water, are simulated. For instance, Zubaidi et al.~\cite{w12071885,w12061628} predict the urban monthly water consumption by learning the temporal dynamics of historical data using neural networks. Hu et al.~\cite{8790060} simulate the urban daily water demand with a CNN and LSTM-based neural network. 

In terms of electricity usage simulation, the goals of different works are at different scales. Li et al.~\cite{LI2019451} studies the city-level electricity consumption during buildings' lifecycle to guide energy conservation policies, while Mahmood et al.~\cite{Mahmood2020SIMULATION} studies at the family level considering family-grained features. Latest researches explicitly consider human daily indoor activities to help predict fine-grained electricity consumption, where the relationship between the usage of electrical appliances and temperature, season, daytime, and family lifestyles is modeled~\cite{schumann:hal-03505242,Peng2012buildings,masciadri2018disseminating}.

For network usage, some works predict network traffic at the cellular tower level based on historical statistics to help cellular operators better allocate the supply and meet the demand, with deep learning techniques such as LSTM~\cite{8057090}, GNN \cite{8466626}, and transfer learning \cite{8667446}. Other works are dedicated to simulating detailed network usage behavior of each individual to support fine-grained network managements~\cite{Xu2021STAN,RING2019156,Hui2022Knowledge}.

\subsection{Social Behavior}
\begin{figure}
    \centering
    \includegraphics[width=0.5\linewidth]{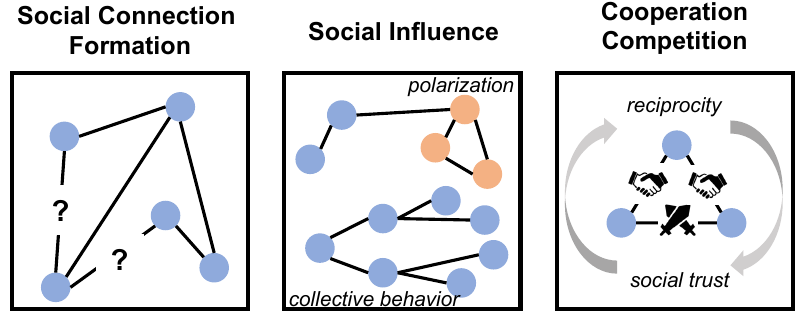}
    \caption{The illustration of social behavior simulation.}
    \label{fig:social}
    \Description[The illustration of social behavior simulation.]{The illustration of social behavior simulation.}
\end{figure}

In recent years, the main focus of social behavior simulation lies in three aspects, including social connection formation simulation, social influence simulation, and cooperation and competition simulation. Their main objectives are making scientific discoveries, such as investigating microscopic social behavior dynamics using agent-based models (ABM)~\cite{bianchi2015agent}. A few researchers are dedicated to improving the simulation accuracy, yet the overall performance still leaves much to be desired. The illustration is shown in Figure~\ref{fig:social}.

\subsubsection{Social Connection Formation}
In general, these researches target deciding whether a specific connection exists in the given partially observed social network, \textit{i.e.}, $P(e_{ij}=1\vert \mathcal{G}')$, where $\mathcal{G}=(V, E)$ is the full network, $V$ and $E$ denote the set of network nodes and edges, respectively. $e_{ij}=1$ denotes there is a connection between node $i$ and $j$, and $\mathcal{G}'$ is partially observed network.

From the global perspective, many works simulate the formation mechanism of real-world social networks with some general properties for the objective of scientific discovery. In other words, expert knowledge of social network formation models $P(\cdot\vert \cdot)$. For example, preferential attachment has been shown as the generic mechanism of scale-free network formation~\cite{barabasi1999emergence}, of which social network is a special case. Triadic closure, its amplification on homophily, and the further cumulative effect on core-periphery structures~\cite{asikainen2020cumulative} are also proved to be able to generate community structures~\cite{bianconi2014triadic}. From the local perspective, link prediction is an important task in social networks. In this research line, $P(\cdot\vert \cdot)$ is replaced with data-driven models, such as neural networks. The accurate modeling of social connections and their evolving dynamics can serve for better decision-making, such as friend recommendation~\cite{geng2015learning}. Many deep-learning-based methods are proposed to beat traditional ones in recent years~\cite{qu2020continuous,pareja2020evolvegcn}, especially based on graph neural networks~\cite{sankar2020dysat}. Some works also leverage prior knowledge in social science for designing neural networks or training loss to achieve higher prediction accuracy of unseen social links~\cite{zhou2018dynamic,wang2023neural}.

\subsubsection{Social Influence}
Besides self-interest as a primary goal, people are also directly influenced by others, like friends or strangers, when making decisions. In this respect, macroscopic behaviors can not be obtained by simply aggregating individual decisions. This kind of tendency usually produces polarized~(\textit{e.g.}, echo chamber) or collective~(\textit{e.g.}, riots) behaviors. Polarization comes up when people prefer local environments similar to themselves, such as neighbors~\cite{schelling1971dynamic,laurie2003role,zhang2004residential}, cultures~\cite{colman1998complexity}, and opinions~\cite{hegselmann2002opinion,baumann2020modeling}. For example, \cite{baumann2020modeling} simulates the evolution of agent opinions and demonstrates the critical role social influence plays in emerging radicalization, which is also consistent with observational data on social media. Collective behaviors indicate emerging events in society with the participation of a large population, both offline and online. This kind of social decision is dominated by others' decisions that people know and see~\cite{granovetter1978threshold}, \textit{i.e.}, social structure, and the influence strength is also dependent on the spatial distance between individuals~\cite{hedstrom1994contagious}. In addition, with the prevalence of online social media, the simulation of collective behaviors on virtual platforms attracts more attention. More factors beyond social influence~(or social structure) are considered to model information cascades in social media, such as timing and user interests~\cite{lu2018collective,cinus2020generating}. As a new modeling paradigm, LLM is increasingly used to simulate the dynamic evolution of people's opinions, emotions, information propagation, interactions and other behaviors in social networks~\cite{gao2023s,acerbi2023large}.

\subsubsection{Cooperation and Competition}
Considering the natural tendency of cooperation and competition among humans, investigating the mechanisms of their emergence is important. Previous research suggests that it can be discovered by the abstract simulation in the context of a prisoner's dilemma. The very first mechanism is reciprocity, which represents the tendency to cooperate with others that may produce higher payoffs. To achieve this goal, human agents can learn from social environments and update their strategies through imitating others'~\cite{cohen2001role,nowak1992tit} or based on their own past experience~\cite{macy2002learning}. In this way, people choose to cooperate with cooperators and defect with defectors. From another perspective, other agents who don't participate in the game directly or the information provided by them can also help to promote cooperation, such as the reputation of circulation among people ~\cite{boero2010bother,conte2002reputation}, punishment on wrongdoers without punishers' additional benefits~\cite{gintis2000strong,bowles2004evolution}. The former provides more information for communication and further finding reliable partners and also serves as a tool against noncooperation behaviors. The latter has a similar function against behaviors of payoff reduction at the population level in a more direct way. The second mechanism is social trust, which means people are willing to cooperate with others they trust. Therefore, identifying trustworthy agents to interact with is a key issue in cooperation emergence. Many studies have investigated potential factors, of which the most important one is social structure~\cite{macy1998evolution,bravo2012trust,fehl2011co}. For example, Bravo et al.~\cite{bravo2012trust} simulate it with laboratory data of real subjects and show that interaction with self-decided agents dynamically can promote local trust and increase cooperation. LLM is also used to simulate the cooperation and competition in various social environment contexts, such as software development~\cite{qian2023communicative}, consumption market~\cite{zhao2023competeai}, \textit{etc}.

\subsection{Economic Behavior}
\begin{figure}
    \centering
    \includegraphics[width=0.5\linewidth]{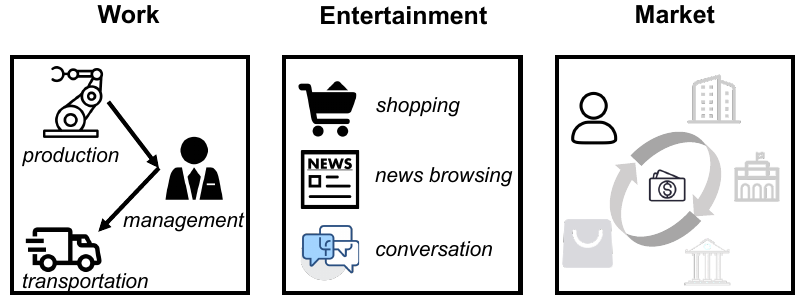}
    \caption{The illustration of economic behavior simulation.}
    \label{fig:economic}
    \Description[The illustration of economic behavior simulation]{The illustration of economic behavior simulation.}
\end{figure}

Although economic behaviors in simulation cover a wide range of real-world scenarios, existing works on simulation concentrate on three primary aspects, i.e., work, entertainment, and market behaviors. In terms of objectives, work, and entertainment are often simulated for optimal decision-making, and market behaviors are typically simulated for scientific discovery. The illustration is shown in Figure~\ref{fig:economic}.

Whether it is decision-making or scientific discovery, the premise of existing economic simulation is to maximize individual or system~(online platform, society, \textit{etc.}) utility. Formally, the optimization goal or simulation assumption is $\max_{a_i, a_s} Utility(a_i, a_s)$, where $a_i$ and $a_s$ denote the actions of individuals and systems, respectively. 
\subsubsection{Work}
Simulation primarily serves as a cost-free decision environment, which is further leveraged for policy or strategy evaluation. In other words, optimization is the essential goal of work simulation. To be more specific, work behaviors cover a wide range of scenarios, including production, management, transportation, etc. In the procedure of production, people make decisions and influence the final outcomes, such as the cumulative error in automotive assembly~\cite{baskaran2019digital} and crop productivity in farm development~\cite{troost2020bioeconomic}. Taking agricultural production as an example, farmers adapt the strategies of land usage, crop allocation, and planting techniques to the climate~\cite{troost2020bioeconomic}. Thus, simulating how people work can facilitate production. In the management simulation, interactions among the population are addressed, such as employee-employer in human resources management~(HRM)~\cite{rachid2018agent,somarathna2020agent}, service personnel-consumer in service operations management~\cite{tian2022hybrid,ajmi2019agent}, etc. There are many types of work in the transportation industry, including driving of car-hailing~\cite{chaudhari2018putting}, logistics~\cite{firdausiyah2019modeling}, and express delivery~\cite{ruvzic2021simulation,chen2017crowdsourced}. In these circumstances, worker actions are evaluated via simulation for the optimization of systemic efficiency and profits.

\subsubsection{Entertainment}
With the rise of the Internet, more and more online entertainment activities have emerged, including online shopping, news browsing, etc., which produces huge economic benefits. As a result, providing satisfactory information for users becomes an essential goal for online services. While data-driven methods have made significant progress in recommender systems~\cite{gao2022survey,zhang2019deepsurvey}, the high cost of testing new algorithms in the real world has become unacceptable. To solve this problem, researchers turn to human behavior simulation, which provides a cost-free environment for algorithm training and testing.

Shi et al.~\cite{shi2019virtual} propose to build a virtual simulation environment for generating customers and their interactions within online shopping platforms. With this environment, the platform can optimize its long-term recommendation policy for improving revenue without the costs of collecting real user behaviors or harming user experiences. Luo et al.~\cite{luo2022mindsim} focus on the scenario of news consumption. Researchers built a simulator based on encode-decoder structure and GAN~\cite{goodfellow2020generative} to generate new users and corresponding click behaviors on recommended news and demonstrated that these simulated data could improve existing recommendation algorithms. Similarly, Zhao et al.~\cite{zhao2021usersim} also propose to generate user behaviors that can be used to train RL algorithms, which achieved performance as well as that of real-world records. From another perspective of algorithm evaluation, Zhang et al.~\cite{zhang2020evaluating} rely on simulated users to assess the performance of the conversational recommender systems in advance of real people using them.

LLM is also used to simulate user behaviors in the context of various online platforms, including user preferences in recommender systems~\cite{wang2023recagent}, users' multi-step searching in website browsing~\cite{gur2023real}, \textit{etc}.

Since there are few works simulating offline entertainment behaviors, we don't include them in this survey.

\subsubsection{Market}
In this subsection, we concentrate on microeconomic and individual behaviors that influence the whole market. Benefited from the rapid development of computational ability, there is growing utilization of agent-based models~(ABM) for microeconomic modeling in markets, which relax the ideal assumptions in conventional methods, such as the hypothesis of rational man, and make further progress in the level of simulation realism~\cite{farmer2009economy,axtell2022agent}.

In terms of research fields, the very first and most important area is price and market. How the price is formed, how the market evolves dynamically, and what the role of human decisions in it are essential questions researchers always concentrate on. The most representative problem is simulating the supply and demand curves driven by agents in the market, including buyers and sellers. Some works attempt to understand what kind of decision mechanisms lead to price equilibrium based on simulation~\cite{gode1993allocative,palmer1994artificial,cliff1998less,albin1992decentralized}. Some researchers further develop general tools for convenient market simulation, which assists in a better understanding of the market with different mechanism designs as well as corresponding outcomes. For example, Byrd et al.~\cite{byrd2020abides} build a simulation system of the financial market upon ABM, with which market dynamics can be reproduced. Based on the proposed simulation environment, Karpe et al.~\cite{karpe2020multi} further investigate optimal trading strategies of order execution with multi-agent reinforcement learning. From another perspective, Amrouni et al.~\cite{amrouni2022ctmstou} focus on the problem of market regimes and demonstrate the importance of regime awareness.

Another prevalent field is studying how humans make decisions when competing with others for optimal utility in markets. One of the important problems is the formation of cooperation under this competitive circumstance. Some works investigate the problem from the scope of interactions in the markets, i.e., who people compete with. With the simulation, researchers can mine essential social processes for cooperation formation~\cite{nowak1992evolutionary,aktipis2004know,aktipis2006recognition,aktipis2011cooperation}. For example, Nowak et al.~\cite{nowak1992evolutionary} show that interacting with immediate neighbors in 2D space enables the cooperation. Others study individual strategies, i.e., how people choose to react. Simulation can be utilized for strategy evaluation and improvements~\cite{bell2001avoiding,bell2003coordination}. In recent years, multi-agent reinforcement learning~(MARL) has been introduced for learning household decision-making, where the utility is optimal objective as the reward~\cite{mi2023taxai,zheng2020ai,curry2022analyzing}. Furthermore, LLM-based agents can be used to simulate the dynamic characteristics of both microeconomics~(such as game theory~\cite{guo2023gpt}) and macroeconomics~\cite{li2023large}.

Note that although behaviors of larger-scale market participants, such as firms, organizations, etc., are also investigated in economics, we do not include them in this survey since we focus more on individual behaviors.

\subsection{Summarization of Current Works}

We summarize the objectives of different types of behavior simulation in Tables~\ref{tab:forSD}, \ref{tab:forDM}. It clearly demonstrates that different research disciplines have different focuses and progress, which shed light on possible directions of future works.

Overall, existing works on cognitive behavior simulation mainly focus on using interpretable models to analyze and understand cognitive behavior, as shown in Tables~\ref{tab:forSD}, \ref{tab:forDM}. Many ideal assumptions and simplifications are made, which leads to low simulation accuracy~\cite{bilewicz2020hate,dykstra2013put}. In particular, they also lack the consideration of modeling individual heterogeneity. With the maturity of natural language processing technology, some researchers suggest that it is time to improve the simulation accuracy for better decision-making~\cite{lopez2021simulating,zhou2020two}. Automatic annotation of opinions based on messages in social networks can yield large-scale real-world datasets, which gives us a unique chance to use data-driven methods for improving simulation accuracy.

In contrast, simulation studies of social, economic, and social behavior are more developed and comprehensive. Both objectives have been addressed. However, we can still find some gaps in the fields. For example, the laws and mechanisms of human resource consumption from an individual perspective remain to be investigated. Research on building realistic simulation environments that support decision-making is still lacking for market behavior.

physiological behaviors refers to those behaviors driven by, or to achieve, or as result of some intrinsic physiological needs. 

\section{Methodology}
\label{sec:method}
The methods used in human behavior simulation can be divided into three categories: 1) Knowledge-driven methods, 2) data-driven methods, and 3) knowledge and data co-driven methods. In the early stages of the research, researchers mainly rely on human experts to derive knowledge from empirical observations and propose physics-based models or heuristic models for human behavior simulation~\cite{helbing1995social,kempe2003maximizing}. Some researchers further summarize domain-specific knowledge and leverage it to give more accurate simulations. These knowledge-based methods are typically robust with good generalizability. However, it is hard for them to simulate human behavior accurately due to its complexity and heterogeneity. With the developments and maturity of the Internet, large-scale human behavior data is collected, and some researchers propose to leverage emerging deep learning techniques for human behavior simulation, which has been an immediate success~\cite{jiang2018deep,wen2021package,yuan2022activity}. These models have achieved high simulation accuracy, yet they are generally less robust compared to knowledge-based models. Many cannot generalize to new application scenarios beyond their training ones. To solve this problem, some recent works suggest combining the two traditional methods together to leverage the advantage of both knowledge-driven and data-driven methods to achieve robust, generalizable, and accurate simulation simultaneously~\cite{Zhang2022PIML}. Although there is no mature paradigm for us to combine the two methods, this direction has drawn a lot of attention and has shown its effectiveness in numerous latest researches~\cite{willard2022integrating}, which demonstrates its great potential research value. 

\begin{figure*}[t]
    \centering
    \includegraphics[width=0.96\textwidth]{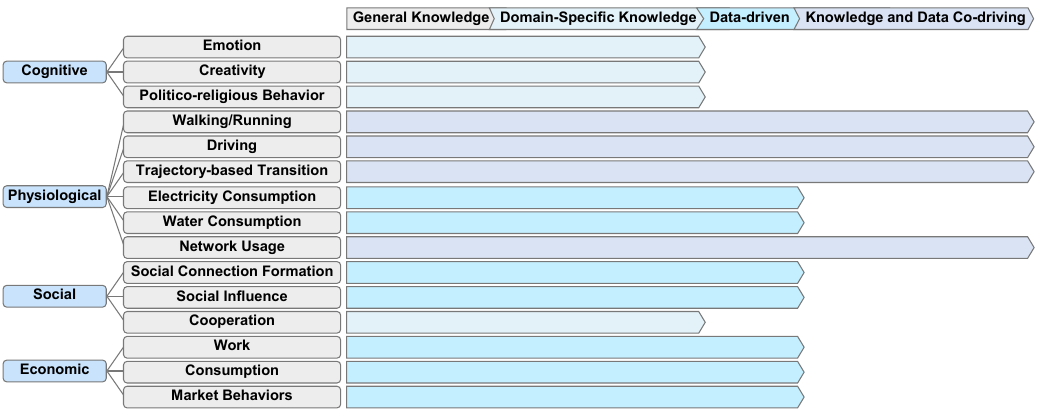}
    \caption{The development status of each research field in terms of methodologies.}
    \label{fig:developmentStatus}
    \Description[The development status of each research field in terms of methodologies.]{The development status of each research field in terms of methodologies.}
\end{figure*}

The current development of each research field is summarized in Figure~\ref{fig:developmentStatus}. Overall, physiological simulation is the most advanced field, where all three methods are widely used. The reason is that human mobility data is one of the most available large-scale data sources, and thereby data-driven methods were developed relatively early. On the contrary, cognitive behavior simulation typically leverages knowledge-driven simulation since it is generally hard to collect corresponding real-world datasets. Social behavior simulation and economic behavior simulation have arrived at the data-driven stage, yet these fields have not found effective methods to combine knowledge-driven and data-driven methods together. 

In this section, we introduce state-of-the-art methods in human behavior simulation divided by the method types. 

\begin{table*}[t]
    \centering
    \small
    \begin{tabular}{m{2cm}<{\centering}  m{2cm}<{\centering}  m{2cm}<{\centering}  m{1.5cm}<{\centering}  m{5.0cm}<{\centering}}
    \toprule
        \multirow{2}*{\textbf{Methodology}} & \multirow{2}*{\textbf{Advantage}}  & \multirow{2}*{\textbf{Disadvantage}} & \multicolumn{2}{c}{\textbf{Examples}} \\
        \cmidrule(lr){4-5}
        & & & \textbf{Model} & \textbf{Features} \\ 
    \midrule
        \multirow{8}{2cm}{Knowledge-driven Model} & \multirow{8}{2cm}{Interpretable, typically highly generalizable} & \multirow{8}{2cm}{Low simulation accuracy and unrealistic} & Social Force & Models and simulates human movement with Newtonian dynamics. \\
        \cmidrule(lr){4-5}
        & & & Percolation Process & Leverages a physical process with phase transitions to simulate diffusions in social media.\\
        \cmidrule(lr){4-5}
        & & & Opinion Dynamics & Leverages domain-specific knowledge in opinion evolution simulation. \\
    \midrule
        \multirow{12}{2cm}{Data-driven Model} & \multirow{12}{2cm}{Relatively high simulation accuracy} & \multirow{12}{2cm}{Uninterpretable, generally has poor performance in generalization.} & RNN & Models human behavior as a sequential process. \\
        \cmidrule(lr){4-5}
        & & & GCN & Simulates human behavior with graph-based modeling. \\
        \cmidrule(lr){4-5}
        & & & GAN & Simulates by generation rather than prediction. \\
        \cmidrule(lr){4-5}
        & & & RL & Does not require labeled data but data dynamically generated during the interactions with the environment. \\
        \cmidrule(lr){4-5}
        & & & GAIL & Simulates by imitating the demonstrations of people. \\ 
    \midrule
        \multirow{4}{2cm}{Knowledge and Data Co-driven Model} & \multirow{4}{2cm}{high simulation accuracy, robust, and interpretable.} & \multirow{4}{2cm}{Difficult to design the model} & PINN & Infuses the knowledge in physical formula form into the neural networks. \\
        \cmidrule(lr){4-5}
        & & & KGENN & Embeds the knowledge in knowledge-graph form into the neural networks. \\
    \bottomrule
    \end{tabular}
    \caption{Comparison among methods used in human behavior simulation.}
    \label{tbl:comparison}
\end{table*}

\subsection{Knowledge-driven Methods}

Knowledge-driven methods typically leverage physical, mathematical, or manually-designed heuristic models for simulation. These types of models are generally interpretable white-box models and have good generalizability. The knowledge used in the models is usually derived from human observations and inductions, which requires a long period of research, and thereby it is also hard for them to simulate human behavior accurately due to its complexity and heterogeneity. 

Knowledge-driven methods are usually the first methods adapted in a new simulation domain, as mentioned above, so the development processes of knowledge-driven methods are of guiding value for a nascent simulation field. We introduce some typical knowledge-based methods and their development process as follows. Note that, in this section, we introduce three representative knowledge-based methods, including the social force model, the percolation process model, and the opinion dynamics model. The details and formulas of the models are ignored as we mainly focus on their main ideas and development processes.

\subsubsection{Social Force Model (SFM)}
\label{sec:SFM}
Proposed by Helbing et al. in 1995 \cite{helbing1995social}, the Social Force Model (SFM) is a classical model that has been widely used in human physical movement simulation until now. SFM treats pedestrians as particles with positions and velocities, and leverages \emph{virtual forces} to model the interactions between pedestrians and the environment (e.g., obstacles or other pedestrians), which further determines their movement trajectories in the scenario. Specifically, consider a room with $N$ pedestrians, where $\mathbf{x}_i(t)\in\mathbb{R}^2$ and $\bm{v}_i(t)\in\mathbb{R}^2$ represent the position and velocity of $i$-th pedestrian at time $t$, SFM assumes that:
\begin{itemize}
    \item A pedestiran wants to reach a certain destination $\bm{d}_i$ as comfortable as possible, that is, s/he wants to move towards it with a desired speed $v_i^{(0)}$.
    \item A pedestrian tends to keep a certain distance from other pedestrians.
\end{itemize}
These two assumptions can be formalized as two forces, including 1) an acceleration force
\begin{equation}
    \bm{f}_i = \frac{1}{\tau}\left(v^{(0)}_i \frac{\bm{d}_i - \bm{x}_i(t)}{\|\bm{d}_i - \bm{x}_i(t)\|} - \bm{v}_i(t)\right),
\end{equation}
and 2) a repulsive force
\begin{equation}
    \bm{g}_{ij} = -\lambda_1 e^{-\lambda_2 \|\bm{x}_j(t) - \bm{x}_i(t)\|}\times\frac{\bm{x}_j(t) - \bm{x}_i(t)}{\|\bm{x}_j(t) - \bm{x}_i(t)\|},
\end{equation}
where $\tau=0.5s$ is a relaxation time and $\lambda_{1,2}>0$ are parameters. With these two forces, the movement of pedestrians can be simulated by
\begin{equation}
    \bm{\dot{x}(t)} = \bm{v}_i(t), \quad \bm{\dot{v}}_i(t) = \bm{f}_i + \sum_{j\neq i} \bm{g}_{ij}.
\end{equation}
By including more types of forces, SFM is able to simulate more complex scenarios. For example, Helbing et al. used a random force $n_i \sim \mathcal{N}(0, \Sigma)$ to model the random drift of pedestrians~\cite{helbing1995social}. They also used an attractive force $g'_{ij}$ to model the effect that pedestrians can sometimes be attracted by other pedestrians (e.g., friends) or objects (e.g., window displays)~\cite{helbing1995social}. In their successive work~\cite{helbing2000simulating}, they leveraged the \emph{herding force} to model the panic effect in a smoke-filled room, where pedestrians tend to move following the average moving direction of the surrounding pedestrians rather than towards the exit. More work has been done to improve the SFM until recently. For example, Dias et al.~\cite{dias2018calibrating} leveraged calibrated SFM to simulate the interaction between pedestrians and vehicles, Li et al.~\cite{li2021modified} used modified SFM to simulate bicycle traffic by introducing the dynamic boundary model and the behavior force model, and Du et al.~\cite{du2022dynamic} introduced the dynamic sensitivity and attention field to simulate the unidirectional pedestrian flow under the effects of individual characteristics and surrounding environments on pedestrian behaviors. However, SFM is still not fully consistent with empirical observations and is sometimes hard to calibrate\cite{moussaid2011simple}, we found a common problem of knowledge-driven methods in human behavior simulation.

\subsubsection{Percolation Theory}
Percolation theory provides a simple yet powerful framework for studying flow and diffusion problems in complex networks, and aids in understanding critical phenomena and phase transitions. Specifically, percolation theory primarily explores how a connected path that traverses the entire system is formed in a medium with a random distribution~\cite{kempe2003maximizing}. In this model, each node can be open or closed, and the whole system can only form a connection when enough nodes are open.
Percolation theory typically uses a percolation threshold ($p_c$) to describe the phenomenon. In a simple two-dimensional lattice model, each lattice site is randomly designated as open (permeable) or closed (impermeable), with a probability of $p$. When $p<p_c$, no continuous path allows fluid to flow through, whereas when $p>p_c$, a connected path forms. The value of $p_c$ varies for different types of networks and lattices. For example, $p_c$ is approximately 0.59 in a two-dimensional square lattice.

Kempe et al.~\cite{kempe2003maximizing} assume that information cascading is a dynamic percolation process and connect social media research to phase transition theories. Based on this assumption, Leskovec et al.~\cite{leskovec2007dynamicsa} study viral marketing. Xie et al.~\cite{xie2021detecting} prove the existence of percolation-like spread in real social media by analyzing many information-spreading trajectories.

However, the percolation model also has its limitations. It assumes that the nodes and edges in the network are uniformly and randomly distributed, which may not correspond to the network structures in the real world. Furthermore, it typically does not consider the time-dependency of network dynamics, which can be important in certain application scenarios.

\subsubsection{Opinion Dynamics}
Opinion dynamics refers to modeling opinion evolution within a population, aiming to understand its dynamic characteristics~\cite{degroot1974reaching}. The model assumes that each individual has their own initial opinion on a set of topics. Before the end of the propagation process, each individual in the model updates their opinion according to specified rules and their current opinion, until the viewpoints of all individuals no longer change. The DeGroot model, proposed in 1974, is one of the earliest and most fundamental models in the field of opinion propagation~\cite{degroot1974reaching}. This model assumes that an individual's opinion in the next iteration is derived from a weighted average of their own current opinion and the opinions of adjacent individuals in their network.
\begin{equation}
    {x_{i}(t+1)=\frac{w_{i i} x_{i}(t)+\sum_{j \in N(i)} w_{i j} x_{j}(t)}{w_{i i}+\sum_{j \in N(i)} w_{i j}}}
\end{equation}
where the weights $w_{i i}$ or $w_{i j}$ represent the individual's confidence in their own opinion or the depth of their friendship with others. Essentially, this model emphasizes the influence of social connections and personal conviction in the evolution of opinions within a group.

Frenchjr et al.~\cite{frenchjr1956formal} introduce opinion dynamics to study the opinion fusion process through interactions among a group of agents. Degroot et al.~\cite{degroot1974reaching} further come up with the DeGroot model assuming that the opinions of each agent tend to be close to the average of its neighbors, and other researchers come up with more opinion dynamics models with different assumptions, such as the Sznajd model, vote model, etc. Based on these models, a lot of macro-level phenomena in social networks are studied. For example, Santos et al.~\cite{santos2021link} use opinion dynamics to simulate how people's opinions become polar. Baumann et al.~\cite{baumann2020modeling} and Cinelli et al.~\cite{cinelli2021echo} simulate the echo chamber effect on social media.

However, opinion dynamics models often rely on simplified assumptions, such as the notion that the patterns of interaction between individuals are fixed or that they ignore the influence of external environmental factors. This can lead to models that fail to accurately reflect the complexity and diversity of the real world. Additionally, in some models, individuals are assumed to exhibit uniformity in behavior and responses in decision-making, which may overlook the individual variability in human behavior.

\subsection{Data-driven Methods}
Data-driven methods are those that learn from real-world data, including statistical models, machine learning models, and deep learning models. Although statistical methods, such as the Bayesian model and decision-tree model, are classic data-driven methods, the use of the recently developed deep learning models in human behavior simulation has grown rapidly in recent years due to their strong representation power. Existing practices have demonstrated that these models are capable of simulating complex human behavior more accurately than knowledge-based ones. They have received a lot of attention and developed fast in recent years. However, they are uninterpretable and generally less generalizable compared to knowledge-based models, and thereby few of these methods are deployed in commercial applications. In this section, we introduce state-of-the-art data-driven models in human behavior simulation.

\subsubsection{Recurrent Neural Network (RNN)}

RNN is a type of neural network designed to process sequential data or time series data. Its main idea is to use the output of the network as a part of the input in the next step to enable the network to give output based on not only the current input but also former inputs. In other words, it is able to learn the complex temporal patterns for simulation. Since human behaviors typically exhibit strong spatial-temporal patterns, many researchers model them as a sequential decision process and leverage RNN to learn the temporal patterns from data for simulation. 

For example, Alahi et al.~\cite{alahi2016socialLSTM} introduce the RNN to simulate human movement by modeling the trajectory of pedestrians as a function of the former trajectories of themselves. Jiang et al.~\cite{jiang2018deep} and Feng et al.~\cite{Feng2020PMF} simulate human transition in urban venues on a coarse-grained scale of space and time in a similar way. Feng et al.~\cite{Feng2018DeepMove} additionally import a historical attention model to capture the multi-level periodicity nature of human mobility behaviors. In the field of human driving behavior, Wang et al.~\cite{wang2018capturing} introduce the RNN to simulate car-following behaviors to enable the drivers to make actions temporally dependently. Sun et al.~\cite{sun2020vehicle} use the RNN to simulate vehicle turning behavior. Suo et al.~\cite{suo2021trafficsim} propose a traffic simulator based on RNN as well. In addition, in economic behavior simulation, Wen et al.~\cite{wen2021package} simulate how couriers pick up packages and use the RNN to consider the spacial-temporal correlations among the representations of packages.

\subsubsection{Graph Convolutional Network (GCN)}

GCN is a recently developed method to process graph-structured data. It takes the adjacency matrix of the graph and the feature of each node as inputs and is good at modeling the structure and attributes of the graph and learning the spatial pattern from data. Researchers typically leverage it to model the relationships and interactions among individuals since human behaviors are influenced by those close to them. For instance, our online posts and retweets are influenced by our friends.

GCN is widely used in physiological behavior simulation and social behavior simulation. For instance, Mohamed et al.~\cite{mohamed2020socialSTGCNN} use the GCN to simulate human movement by modeling it as a graph where edges exist between each pedestrian and viewable nearby pedestrians. Wang et al.~\cite{wang2019origin} combine the GCN and RNN to simulate human mobility focusing on spatial and temporal information, respectively. Suo et al.~\cite{suo2021trafficsim} and Zhang et al.~\cite{zhang2022trajgen} use GCN to simulate human driving behavior as well in a similar way. Sankar et al.~\cite{sankar2020dysat} leverage GCN to simulate social network formation and online diffusion, which achieves state-of-the-art performance.

\subsubsection{Generative Adversarial Network (GAN)} 

GAN is a recently innovated generative model that introduces the concept of adversarial learning. It trains two models simultaneously: the generator and discriminator. The generator generates samples with random noise as seeds, and the discriminator discriminates whether the input samples are real-world data or data generated by the generator. The training process of the GAN is a two-player zero-sum minimax game between the discriminator and the generator. By training them simultaneously, the generator can generate samples so close to real-world data distribution that even a well-trained discriminator is confused.

Since GAN has the ability to model the uncertainty by inferring the data distribution, it usually has two roles in human behavior simulation. One is modeling the heterogeneity of human behaviors. For example, Hui et al.~\cite{Hui2022Knowledge} simulate the heterogeneous network usage behavior with GAN. Shi et al.~\cite{shi2019virtual} use the GAN to generate heterogeneous customers and simulate how they interact with the recommender system, which performs better than traditional supervised learning approaches in online testing. The other is directly formulating simulation as a generation problem. For example, Gupta et al.~\cite{gupta2018social}, and Feng et al.~\cite{feng2020learning} simulate human movement behavior with the GAN to represent the inherently multimodel in the motion of pedestrians. Zhang et al.~\cite{zhang2022systematic} simulate driving behaviors with the GAN for a similar reason. However, the performance of GAN is generally less stable than supervised learning methods.

\subsubsection{Reinforcement Learning (RL)} 

RL is a powerful technique in modern deep learning. It introduces another machine learning paradigm alongside Supervised Learning (SL). Unlike the SL, RL does not require labeled real-world data but learns by data from a trial and error process. Specifically, it trains an intelligent agent to choose the best actions according to its current state by rewarding desired behaviors in the learning process. Compared with other data-driven methods, RL has fewer applications in human behavior simulation since the simulation performance is dependent on the reward function we define. However, a good reward function is generally hard to design because the motivations of most human behaviors are unclear.

Some researchers applied RL in physiological and economic behavior simulation. For instance, Luo et al.~\cite{Luo2022RLMob} predict successive mobility behaviors by modeling the user's demands as a discrete-continuous hybrid function and optimizing it with RL. Zhang et al.~\cite{zhang2022trajgen} combine the RL and SL to simulate driving behaviors. It first predicts the trajectories of the vehicles by the SL and then adjusts the trajectories to avoid collisions by RL. Lefebvre et al.~\cite{lefebvre2018contrasting} leverage an RL-based model to simulate and explain the emergence of money, which achieves satisfactory performance.

\subsubsection{Generative Adversarial Imitation Learning (GAIL)}

GAIL can be seen as a combination of \emph{Inverse Reinforcement Learning (IRL)} and RL. IRL is the inverse version of the RL: RL optimizes a policy under the guidance of a reward function, while IRL learns a reward function through a demonstration policy from experts. By combining these two processes, In this way, GAIL can learn to imitate the experts' behavior, which inspired researchers to apply it in human behavior simulation. Specifically, in many cases, we can obtain human behavior data, which acts as expert demonstrations and thus makes GAIL naturally fit for simulating human behavior. 

What makes it better than other IL methods is that it does not directly imitate the expert policy but obtains the inherit reward function first and obtains the imitation policy through conducting RL on it, which has better generalization ability as a result. GAIL is suitable for problems that model human behavior as sequential processes. For example, Yuan et al.~\cite{yuan2022activity} simulate human movements with their activities as a sequential process and generate the activity trajectories with a GAIL-based model. Zhang et al.~\cite{zhang2019unveiling} model how taxi drivers seek passengers as a sequential process and simulate it with GAIL as well.

\subsection{Knowledge and Data Co-driven Methods}

The knowledge and data co-driven methods come from the idea that the advantages of knowledge-driven and data-driven methods can complement each other. Specifically, knowledge-driven methods are typically more generalizable than data-driven methods, while their simulation accuracy is lower than data-driven ones. Thus, combining the two methods can potentially achieve robust, generalizable, and accurate simulation at the same time. 

Apart from its great research value, how to effectively combine these two methods in human behavior simulation is still an emerging direction that remains to be explored. In this section, we introduce two types of state-of-the-art attempts that successfully combine the two methods together and have been proven to be useful in human behavior simulation.

\subsubsection{Deep Learning aided Knowledge Discovery}
In recent years, integrating data-driven methods into various traditional scientific fields to address increasingly complex problems has gradually become a mainstream trend. In this process, deep learning-aided knowledge discovery has demonstrated significant potential for accelerating computational speed and advancing knowledge discovery, especially in areas where traditional hypothesis-driven methods are limited.

Deep learning is widely used to handle vast amounts of data, effectively accelerating traditional knowledge-based methods. For example, Wang et al.~\cite{wang2019massive} propose a framework that trains neural networks with simulated data from mechanical models to explore extensive parameter spaces more efficiently. Additionally, Merchant et al. ~\cite{merchant2023scaling} applied graph neural networks to predict new stable materials, where AI successfully predicted 2.2 million unique crystal structures, equating to around 800 years of human research in materials science.

Deep learning can also enable the discovery of new knowledge or patterns, which might be difficult to identify using traditional knowledge-driven methods.
For example, Petersen et al.~\cite{petersen2019deep} propose a deep symbolic regression method for recovering mathematical expressions from data via risk-seeking policy gradients. Petersen et al.~\cite{zhang2023deep} propose a novel neural network architecture for extending symbolic regression to parametric systems. This neural network-based architecture is also combined with other deep learning frameworks, such as convolutional neural networks, to analyze varying spring systems in one-dimensional images.

Deep learning has shown tremendous potential for aiding knowledge discovery in scientific fields such as biology, materials science, and medicine. However, the application of deep learning to assist knowledge discovery in human behavior simulation is still relatively rare. This paradigm holds the promise of uncovering the underlying mechanisms driving human behavior and enhancing the interpretability of human behavior models. Deep learning-aided knowledge discovery can potentially advance research in human behavior simulation further.

\subsubsection{Data-driven Methods Advanced by Knowledge}~~~~~~

\noindent\textbf{Knowledge as Observational biases.}
Observational bias is considered the simplest way to directly incorporate knowledge into a data-driven model, by training the model with carefully collected, crafted, or augmented data that embodies the underlying physics or structures which dictate their generation\cite{karniadakis2021physics}. 
We find, in the field of human behavior simulation, several works have been feeding their data-driven neural networks with extra data generated by knowledge-based mechanical models as well as the raw collected observations. A crowd movement simulation work \cite{zhang2022physics} implements a physics-infused machine learning framework where a teacher-forcing module is pre-trained on the data generated by the classical Social Force Model before a student-forcing module is fine-tuned on real observations. A vehicle trajectory generation work\cite{zhang2022trajgen} simulates the driving behaviors with RL, which is trained on data generated by a neural network that is explicitly constrained by prior collision or off-road avoidance rules.
There are also several works augmenting the input data of their neural networks by constructing extra data features based on prior knowledge. Specifically, a crowd movement simulation work\cite{SONG2018pedestrianANN} not only feeds the positions and speeds of the target agent and its neighbors into the neural network, but also calculates the desired moving directions of the agents using the classical pathfinding algorithm \textit{A$^*$} \cite{hart1968formal} as extra features to be considered by the neural network. A transition trajectory generation work\cite{feng2020learning} augmented the input data by constructing a transition probability matrix between the pairs of locations as extra features, via the statistics on the dataset, which enables the neural network to consider the prior popularity of the trip's origin and destination.

\vspace{1em}
\noindent\textbf{Knowledge as Inductive Biases.}
Inductive bias is another typical way to incorporate knowledge into a data-driven method, such as an artificial neural network. Specifically, it means designing a tailored model architecture to guarantee its output simplicity satisfies a set of given rules\cite{karniadakis2021physics}. A typical example is Convolutional Neural Networks (CNN), which leverage sliding convolutional kernels to ensure the Spatial translation invariance that widely exists in images\cite{pun2019physically}. In the human behavior simulation field, inductive bias is widely used to incorporate all kinds of empirical or physical laws into neural networks.

Wang et al. \cite{wang2023neural} proposed a Neural Common Neighbor method to predict social connection formation by advancing traditional common neighbor-based methods with neural networks. These methods model social relationships as a graph and predict the missing connection between any two nodes $i$ and $j$ by a score function that counts their common neighbors $N(i)\cap N(j)$, i.e. $S(i,j)=\sum_{u\in N(i)\cap N(j)} 1$. While Wang et al. replace the summing term with the output of a neural network and extend the summing range from direct neighbors to high-order neighbors.

In the realm of physical movement, a well-established empirical law is that pedestrians tend to pay greater attention to individuals nearby rather than those at a distance since the former are more likely to affect their movement. Consequently, research in this domain \cite{alahi2016socialLSTM,gupta2018social,mohamed2020socialSTGCNN,zhang2022physics} often leverages a well-designed pooling mechanism to incorporate this law. For example, Alahi et al. introduced a ``social pooling'' mechanism, which enables the exchange of hidden-layer information among pedestrians that are spatially close to each other, and thus filters the information from distant individuals. On top of this social law, Zhang et al. \cite{zhang2022physics} further utilized the physical law of pedestrian movement. Specifically, rather than directly predict the movement trajectories, they leveraged a graph neural network to predict the ``repulsive force'' term, which is used to represent the interaction between pedestrians in the social force model (see Section~\ref{sec:SFM}), and thus ensure that the generated movement conforms to Newton's laws of mechanics.

\vspace{1em}
\noindent\textbf{Knowledge as Learning Biases.}
In such knowledge and data co-driven methods, knowledge is typically introduced into the model training in the form of \textit{assumptions}. This includes its use in designing multi-task loss functions and novel training algorithms. Another relatively distinct approach to integration involves using knowledge to refine the outcomes of data-driven models. This is commonly seen in simulations of physiological behaviors, which must strictly adhere to physical constraints.

In the simulation of physiological behavior, the knowledge used for designing loss functions often relates to human psychological states, social attributes, or behavioral patterns. For instance, Yao \textit{et al.}~\cite{yao2021pedestriancross} designed a loss function modeling psychological decision-making for multi-task learning in predicting human pedestrian crossing behaviors. In traffic simulation, Peng~\textit{et al.}~\cite{NEURIPS2021_594ca7ad} demonstrated modeling of vehicle drivers' trade-offs between personal and collective benefits. Another work~\cite{Suo_2021_CVPR} incorporated human collision avoidance as common knowledge into the loss function. Feng~\textit{et al.}~\cite{feng2020learning} considered the temporal periodicity and spatial continuity of human movement behaviors as constraints to enhance the realism of simulations of human mobility data.
Another way to leverage knowledge is through the modification of results obtained from data-driven methods, commonly used in simulations of autonomous driving. This includes the refinement of automatic lane-changing through mechanistic models~\cite{zhu2021combined}, the adjustment of vehicle turning behaviors based on physical constraints~\cite{sun2020vehicle}, and the correction of generated trajectories for collision avoidance~\cite{zhang2022trajgen}.

In social and economic behavior simulations, comprehensive or widely recognized domain knowledge can be directly used to guide the learning of data-driven models. For example, some work on link prediction (including in social networks) integrates the assumptions of triadic closure and social homophily into the learning of node representations~\cite{zhou2018dynamic}. Okawa \textit{et al.}~\cite{okawa2022predicting} incorporates sociological knowledge of opinion dynamics represented by ODE and introduces it into the model as an auxiliary task for predicting opinion dynamics. In macroeconomic simulations, RL has shown promising results in the optimal taxation problem, where the optimization objectives are derived from economists' precise definitions of utility functions for households and governments~\cite{mi2023taxai,zheng2020ai,curry2022analyzing}.

\subsubsection{Large Language Model}
Supported by massive data training, large language models (LLMs) have not only acquired general knowledge but also developed unprecedented human-like characteristics. This makes the simulation of LLM-based agents a popular research direction. Although current related research on human behavior simulation covers a wide range of fields, most studies adhere to the basic paradigm involving LLM agents, environments, and their interactions. To simulate more realistic and intelligent agent behavior, LLMs need to possess capabilities for environmental perception, reasoning, and decision-making, as well as the ability to play heterogeneous roles~\cite{gao2023large}.

In terms of interaction with the environment, text is a widely adopted format, wherein LLMs observe and understand the impact of agent behaviors within an environment through textual information. For some more complex environments that cannot be fully described by text alone, some research efforts have provided tools for LLMs to facilitate the perception of such environments. The environment perception relies on the common sense/knowledge inherent in LLMs.

Reasoning and decision-making are key for LLM agents to complete tasks that are complex and span over longer simulation steps. The current agent design usually includes modules for planning and memory, which are used to break down complex tasks into simpler ones and to store crucial information as references for future decisions, respectively. In addition to storing factual information such as observations of the environment, the memory module also forms higher-level experiential summaries through reflection to assist future strategic behavior. Among these modules, the planning ability stems from the combination of LLM’s common sense and external knowledge, while the capabilities for memory and reflection benefit from the reasoning and mining of feedback data~\cite{zhu2023ghost,shinn2023reflexion}.

A salient characteristic of human behavior is its heterogeneity, which poses a challenge for accurate and realistic simulation. Existing studies have confirmed that LLMs can exhibit behaviors with distinct personality traits under mechanisms of prompt engineering or tuning. For instance, AI Town demonstrates diversified social activities in LLM agents by setting different profiles~\cite{park2023generative}. Here, prompt engineering utilizes LLM’s common sense/knowledge about characteristics of people in the real world, while tuning requires extensive domain-specific data.

\section{Trends and Future Directions}

Human behavior simulation is becoming an increasingly indispensable topic in research. Although extensive work has been accomplished, there are still many directions worth exploring. In Tables~\ref{tab:forSD} and \ref{tab:forDM}, we summarize the objectives that researchers focus on in different types of behavior simulation research. It can be clearly seen that many boxes are still blank, which indicates future directions. For example, many studies still lack attention to microscopic behavioral simulations. Beyond the future directions of each behavior, which we have already discussed in the previous section, we have also summarized the open problems that we think are the most valuable below.

\noindent
\textbf{Multi-behavior joint Simulation}.
Existing research typically focuses on a single behavior, such as walking or driving. However, human behaviors are very complex. Single-behavior simulation cannot support many real-world applications well. For instance, if we want to optimize the operating efficiency and cost of a factory, the simulation of various behaviors of workers is necessary. From another perspective, jointly simulating multiple behaviors can be a promising direction for improving the accuracy of the simulation. The reason is that many human behaviors are inextricably linked. If we can successfully model such relationships, we can introduce additional information to the simulation and raise the simulation accuracy. This point has been proved in recommender systems, and we suppose it to be a good opportunity for simulation.

\noindent
\textbf{Multi-scale Behavior Simulation}.
Multi-scale simulation is a common concept that originated from physics simulation, which refers to the modeling of multiple scales of a system. The core motivation of it is that in many cases, single-scale models have their own defects and cannot solve all problems. Macro models have high computational efficiency, but their granularity is too coarse to meet the needs of many microscopic applications. On the contrary, micro-scale simulation has fine-grained granularity but often requires a very large amount of calculation, and the efficiency becomes a bottleneck of application. In human behavior simulation, we believe that multi-scale simulation is also important. On the one hand, the complexity of a person is much higher than that of a particle. Thus, if we do fine-grained modeling of each person, it can certainly not be extended to applications at the urban scale such as urban security management. Therefore, multi-scale joint simulation is very important to support flexible practical applications. On the other hand, modeling the internal connection between macro and micro human behaviors can help us achieve more accurate simulations. Due to data privacy and security issues, macroscopic data aggregated from individuals are persuasive, while individual data are scarce. Thus, how to model the connection between macro and micro to jointly leverage both data sources to realize accurate simulation is an important open problem. Some researchers have made some preliminary progress in this direction. For example, Yang et al.~\cite{yang2020devil} propose an optimization framework to leverage both macroscopic and microscopic data to predict macroscopic statistics. Vlachas et al.~\cite{vlachas2022multiscale} leverages an autoencoder to model the connection between fine and coarse-grained dynamics, which helps them successfully simulate the multi-scale dynamics of diverse complex systems.

\noindent
\textbf{Knowledge and Data Co-driven Simulation}.
As illustrated in Section~\ref{sec:method}, the most commonly used methods in the field of human behavior simulation are knowledge-driven methods and data-driven methods. However, both of them have their own problems. Knowledge-driven methods, such as the social force model inspired by physics, are robust, yet typically do not have enough representation capability to accurately model complex human behaviors. Data-driven methods excel at modeling high-dimensional data and are thus capable of simulating complex and heterogeneous human behaviors. However, it typically has poor generalizability. The deep learning models for human behavior simulation are black boxes and often violate physical constraints or common sense. How to leverage the strength of both types of methods to avoid their shortcomings at the same time to achieve accurate and generalizable simulation is an important and promising direction. Researchers have made some preliminary attempts. For instance, Zhang et al.~\cite{Zhang2022PIML} propose a physics-infused machine learning framework to leverage the strength of both the data-driven model and physics-based model for crowd simulation. Nevertheless, this method is limited to the knowledge of a specific physics form. How to further integrate more knowledge with data seamlessly for human behavior simulation is an important future direction.

\noindent
\textbf{Behavior Simulation Driven by LLM Agents}.
LLM agents have become a promising research direction in terms of simulation and decision-making in recent years, especially in the scenarios of human behavior. The core advantages of LLM-empowered agents lie in their sufficient intelligence and human-like characteristics that enable automated, adaptive behaviors and feedback within environments. Moreover, the heterogeneity of multi-agent simulations can be realized through simple profile prompts or tuning. In contrast, traditional rule-based (knowledge-driven) or neural network (data-driven or knowledge and data co-driven) agents require extensive expert knowledge or large-scale data collection and training. However, large language models also need to address efficiency issues to elevate the number of simulated agents to a higher level. The robustness of language generation is another pressing issue that needs resolution~\cite{gao2023large}.

\noindent
\textbf{Behavior Simulation in Abnormal Conditions}.
Applications for abnormal scenarios are often indispensable, and their value may even be higher than normal scenarios. For example, the simulation of residents' behaviors in disasters or sudden accidents is crucial to building architecture design and emergency management. In these scenarios, human behaviors are often significantly different from normal conditions due to the effects of emotions such as panic and anxiety. However, due to the scarcity of abnormal scenarios and the difficulty of data collection, existing methods, especially those deep learning ones, are typically ineffective. How to accurately simulate human behaviors in abnormal scenarios remains to be an open problem.

\section{Conclusion}
In this study, we present an extensive review of the most notable works on human behavior simulation. We introduced the basic concepts and summarized the problems, objectives, and commonly used methodologies for human behavior simulation. We highlighted the differences among the research advancements in different disciplines, which points to promising future directions. Finally, we discussed the research trends, challenges, and open problems, which reveal the gaps and opportunities for high-impact research. We hope this survey can provide researchers and practitioners with a comprehensive view of the area and stimulate more related research.

\bibliographystyle{unsrt}
\bibliography{ref}

\end{document}